\documentclass[twocolumn,superscriptaddress,amsmath,amssymb,floatfix,aps,pre,showpacs]{revtex4}
\usepackage{amsmath}    % need for subequations
\usepackage{graphicx}   % need for figures
\usepackage{verbatim}   % useful for program listings
\usepackage{color}      % use if color is used in text
\usepackage{subfigure}  % use for side-by-side figures
\usepackage{hyperref}   % use for hypertext links, including those to external documents and URLs
\usepackage{mathrsfs}
\usepackage{amssymb,amsfonts,amsmath}    % need for subequations
\usepackage{graphicx}   % need for figures
\usepackage{verbatim}   % useful for program listings
\usepackage{color}      % use if color is used in text
\usepackage{subfigure}  % use for side-by-side figures
\usepackage{hyphenat}
\usepackage[normalem]{ulem}
\usepackage{array}
\usepackage{epstopdf}

\usepackage{color}      % use if color is used in text
\usepackage{soul}

\newcommand{\fref}[1]{Fig.~\ref{fig:#1}}

\newcommand{\flabel}[1]{\label{fig:#1}}
\newcommand{\eref}[1]{Eq.~\ref{eqn:#1}}

\newcommand{\erefstwo}[2]{Eqs.~\ref{eqn:#1}~and~\ref{eqn:#2}}

\newcommand{\elabel}[1]{\label{eqn:#1}}

\newcommand{\br}[1]{\ensuremath{\left ( #1\right )}}

\newcommand{\sqbr}[1]{\ensuremath{\left [ #1\right ]}}

\newcommand{\sqbrin}[1]{\ensuremath{[#1]}}
\newcommand{\brin}[1]{\ensuremath{(#1)}}
\newcommand{\avgin}[1]{\ensuremath{\langle#1\rangle}}
\newcommand{\avg}[1]{\ensuremath{\left < #1 \right >}}
\newcommand*{\unit}[1]{\ensuremath{\mathrm{\,#1}}}

\begin{document}
\bibliographystyle{prsty} % Choose Phys. Rev. style for bibliography

\title{Fundamental Limits to Cellular Sensing} 
\author{Pieter Rein ten Wolde} \affiliation{FOM
  Institute AMOLF, Science Park 104, 1098 XE Amsterdam, The
  Netherlands} 
\author{Nils B. Becker} \affiliation{Bioquant, Universtit\"at Heidelberg, Im
Neuenheimer Feld 267, 69120 Heidelberg, Germany}
\author{Thomas E. Ouldridge}\affiliation{Department of Mathematics, Imperial
  College, Queeen's Gate, London, SW7 2AZ, UK}
\author{Andrew Mugler} \affiliation{Department of Physics, Purdue University, West Lafayette, IN
        47907, USA}

\begin{abstract}
  In recent years experiments have demonstrated that living cells can
  measure low chemical concentrations with high precision, and much
  progress has been made in understanding what sets the fundamental
  limit to the precision of chemical sensing.  Chemical concentration
  measurements start with the binding of ligand molecules
  to receptor proteins, which is an inherently noisy process,
  especially at low concentrations.  The
  signaling networks that 
  transmit the information on the ligand concentration from the
  receptors into the cell have to filter this
  noise extrinsic to the cell as much as possible. These networks,
  however, are also stochastic in nature, which means that they will
  also add noise to the transmitted signal.  In this review, we will
  first discuss how the diffusive transport and binding of ligand to
  the receptor sets the receptor correlation time, and then how
  downstream signaling pathways integrate the noise in the receptor
  state; we will discuss how the number of receptors, the receptor
  correlation time, and the effective integration time together set a
  fundamental limit on the precision of sensing. We then discuss how
  cells can remove the receptor noise while simultaneously suppressing
  the intrinsic noise in the signaling network. We describe why this
  mechanism of time integration requires three classes of
  resources---receptors and their integration time, readout molecules,
  energy---and how each resource class sets a fundamental sensing
  limit. We also briefly discuss the scheme of maximum-likelihood
  estimation, the role of receptor cooperativity, and how cellular
  copy protocols differ from canonical copy protocols typically
  considered in the computational literature, explaining why cellular
  sensing systems can never reach the Landauer limit on the optimal
  trade-off between accuracy and energetic cost.
\end{abstract}

\maketitle

\section{Introduction}
 Living cells can sense changes in their environment
 with extraordinary precision. Receptors in our visual
system can detect single photons \cite{Rieke1998}, some animals can
smell single molecules \cite{Boeckh1965}, swimming bacteria can
respond to the binding and unbinding of only a limited number of
molecules \cite{berg1977,Sourjik:2002fk}, and eukaryotic cells can
respond to a difference in $\sim 10$ molecules between the front and
the back of the cell \cite{Ueda:2007uq}. Recent experiments suggest
that the precision of the embryonic development of the fruitfly {\it
  Drosophila} is close to the limit set by the available number of
regulatory proteins \cite{Gregor2007,Erdmann2009,Dubuis:2013cp}. This
raises the question what is the fundamental limit to the precision of
chemical concentration measurements.

Living cells measure chemical concentrations via receptor proteins,
which can either be at the cell surface or inside the cell.  These
measurements are inevitably corrupted by two sources of noise. One is
the stochastic transport of the ligand molecules to the receptor via
diffusion; the other is the binding of the ligand molecules to the
receptor after they have arrived at its surface. Berg
and Purcell pointed out in the seventies that cells can reduce
the sensing error by increasing the number of measurements, and that
they can do so in two principal ways \cite{berg1977}. The first is to
simply increase the number of receptors. The other is to increase the
number of measurements per receptor. In the latter scenario, the cell
infers the ligand concentration not from the instantaneous
ligand-binding state of the receptor, but rather from its average over
some integration time $T$. This time integration has to be performed
by the signaling network downstream of the receptor proteins.

In recent years, tremendous progress has been made in understanding
how accurately cells can measure chemical concentrations
\cite{berg1977,Ueda:2007uq,bialek2005,levinepre2007,levineprl2008,wingreen2009,levineprl2010,mora2010,Govern2012,Mehta2012,Skoge:2011gi,Skoge:2013fq,Kaizu:2014eb,Govern:2014ef,Govern:2014ez,Lang:2014ir}. Most
of these studies assume that the cell estimates the concentration via
the mechanism of time integration as envisioned by Berg and Purcell
\cite{berg1977,Ueda:2007uq,bialek2005,levinepre2007,levineprl2008,levineprl2010,Govern2012,Mehta2012,Skoge:2011gi,Skoge:2013fq,Kaizu:2014eb,Govern:2014ef,Govern:2014ez},
although Mora, Endres, Wingreen and others have shown that under
certain conditions a better estimate of the concentration can be
obtained via maximum likelihood estimation
\cite{wingreen2009,mora2010,Lang:2014ir}. In this review, we will
limit ourselves to sensing static concentrations, which do not change
on the timescale of the response, and we will focus on the mechanism
of time integration, although we will also briefly discuss the scheme
of maximum likelihood estimation. This review will follow a series of
papers written by the authors, but, in doing so, will also discuss
other relevant papers.

Specifically, in this review we will address the following questions:
if the downstream signaling network integrates the state of the
receptor over some given integration time $T$, what is then the
sensing error? This is the question that was first addressed by Berg
and Purcell \cite{berg1977}, and later followed up by many authors
\cite{Ueda:2007uq,bialek2005,levinepre2007,levineprl2008,levineprl2010,Skoge:2011gi,Skoge:2013fq,Kaizu:2014eb}. The
answer depends on the correlation time of the receptor, which is
determined by the stochastic arrival of the ligand molecules at the
receptor by diffusion and on the stochastic binding of the ligand
molecules to the receptor. Recently, the correct expression for the
correlation time and hence the sensing error has become the subject of
debate \cite{berg1977,bialek2005,Kaizu:2014eb}, which we will review
in section \ref{sec:BP}. The next question is: How do signaling
networks integrate the receptor state? Do they integrate it uniformly
in time, as assumed by Berg and Purcell? If not, can cellular sensing
systems then actually reach the sensing limit of Berg and Purcell? As
we will see, signaling networks can not only reach the Berg-Purcell
limit, but, in some cases, even beat it by 12\%
\cite{Govern2012}. 

Importantly, the signaling network downstream of the receptor is
stochastic in nature, which means that while the network is removing
the extrinsic noise in the receptor signal, it will also add its own
intrinsic noise to the transmitted signal. Most studies have ignored
this intrinsic noise in the signaling network, essentially assuming
that it can be made arbitrarily small
\cite{berg1977,Ueda:2007uq,bialek2005,levinepre2007,
  levineprl2008,wingreen2009,endres2010,levineprl2010,Govern2012,Skoge:2011gi,Skoge:2013fq,Berezhkovskii:2013hq,Kaizu:2014eb,Lang:2014ir}. However,
can signaling networks remove the extrinsic noise in the input signal
and simultaneously suppress the intrinsic noise of the signaling
network \cite{Govern:2014ef,Govern:2014ez}? If so, what
resources---receptors, time, readout molecules, energy---are required?
Do these resources fundamentally limit sensing, like weak links in a
chain? Or can they compensate each other, leading to trade-offs
between them?  We will see that equilibrium networks, which are not
driven out of thermodynamic equilibrium, can sense---energy
dissipation is not essential for sensing
\cite{Govern:2014ez}. However, their sensing accuracy is limited by
the number of receptors; adding a downstream network can never improve
the precision of sensing.  This is because equilibrium sensing systems
face a fundamental trade-off between the removal of extrinsic and
intrinsic noise \cite{Govern:2014ez}. Only non-equilibrium systems can
lift this trade-off: they can integrate the receptor state over time
while suppressing the intrinsic noise by using energy to store the
receptor state into stable chemical modification states of the readout
molecules \cite{Mehta2012,Govern:2014ef,Govern:2014ez}. Storing the
state of the bound receptor over time using a canonical push-pull
signaling network requires at least one readout molecule to store the
state and at least $4 k_BT$ of energy to store it reliably
\cite{Govern:2014ef}. Non-equilibrium systems thus require three
resource classes that are fundamentally required for
sensing---receptors and their integration time, readout molecules, and
energy. Each resource class sets a fundamental sensing limit, which means that
the sensing precision is bounded by the limiting resource class and
cannot be enhanced by increasing another class.

Last but not least, we will address the question of whether cellular
sensing involves computations that can be understood using ideas from
the thermodynamics of computation
\cite{Bennett:1982hi,Landauer1961}. Cells seem to copy the
  ligand-binding state of the receptor into chemical modification
  states of downstream readout molecules
\cite{Mehta2012,Govern:2014ef,Govern:2014ez}, but can this process be
rigorously mapped onto computational protocols typically considered in
the computational literature \cite{Ouldridge:2015vi}? If so, how do
these cellular copy protocols compare to thermodynamically optimal
protocols? Can they reach the Landauer bound, which states that the
fundamental limit on the energetic cost of an
irreversible computation is $k_B
T \ln(2)$ per bit?  We will see that cellular copy operations differ
fundamentally in their design from thermodynamically optimal
protocols, and that as a result they can never reach the Landauer
limit, regardless of parameters \cite{Ouldridge:2015vi}.

\section{The Berg-Purcell limit}
\label{sec:BP}
\subsection{Set up of the problem}
Berg and Purcell and subsequent authors
\cite{berg1977,Ueda:2007uq,bialek2005,levinepre2007,levineprl2008,levineprl2010,Skoge:2011gi,Skoge:2013fq,Berezhkovskii:2013hq,Kaizu:2014eb}
considered the scenario in which the cell estimates the ligand
concentration $c$, assumed to be constant on the timescale of the
response, by monitoring the occupancy of the receptor to which
the ligand molecules bind and unbind. The key idea is that the
cell infers the concentration by estimating the true average
receptor occupancy $\overline{n}$ from the average occupancy ${n}_T$
over some integration time $T$, and by inverting the input-output
relation $\overline{n}(c)$ \cite{berg1977}. 
A central result is that for a single receptor.
 The time average of its occupancy $n(t)$ over the
integration time $T$ is $n_T = (1/T) \int_0^T n(t^\prime) dt^\prime$.
Using error propagation, the fractional error in the estimate of the
concentration, $\delta c / c$, is then given by
\begin{equation}
\left(\frac{\delta
    c}{c}\right)^2=\frac{1}{c^2}\left(\frac{dc}{d\overline{n}}\right)^2
\sigma_{n_T}^2,
\elabel{error}
\end{equation}
where $\sigma_{n_T}$ is the variance in the time-averaged
occupancy $n_T$, and $d\overline{n}/dc$ is the gain, which determines
how the error in the estimate of $\overline{n}$ propagates to that in
the estimate of $c$. The gain can be obtained from the input-output
relation $\overline{n}(c)=c / (c+K_D)$, where $K_D$ is the
receptor-ligand binding affinity: $d\overline{n}/dc=\overline{n}(1-\overline{n})/c$. In the limit that the integration
time $T$ is much longer than the receptor correlation time $\tau_c$,
the variance in the estimate $n_T$ of the true mean occupancy
$\overline{n}$ is 
\begin{align}
\sigma_{n_T}^2&\approx\frac{2\sigma^2_n \tau_c}{T}=\frac{P_n\br{\omega=0}}{T}=
\frac{2{\rm     Re}\sqbr{\widehat{C}_n\br{s=0}}}{T},\elabel{sigma_nT}
\end{align}
where  the instantaneous variance
$\sigma^2_n=\avg{n^2}-\avg{n}^2=\overline{n}\brin{1-\overline{n}}=p(1-p)$,
with $p=\overline{n}$ the probability that the receptor is ligand
bound, and
$P_n\brin{\omega}$ and $\widehat{C}_n\brin{s}$ are respectively the power spectrum
and the Laplace transform of the correlation function $C_n\brin{t}$ of
$n(t)$. The above expression shows that the variance in the average
$n_T$ is given by the instantaneous variance $\sigma^2_n$ divided
by $T/(2 \tau_c)$, which can be interpreted as the number of independent measurements of
$n(t)$. Inserting \eref{sigma_nT} into \eref{error} yields
\begin{equation}
\left(\frac{\delta c}{c}\right)^2=\frac{2 \tau_c}{p(1-p) T}.
\elabel{error2}
\end{equation}
This is indeed the sensing error based on $T/(2\tau_c)$ independent
concentration measurements.

\eref{error2} holds for any
single receptor, be it a promoter on the DNA, a receptor on
the cell membrane, or a receptor protein freely diffusing inside the
cytoplasm \cite{Paijmans:2014cb}. All we need to do to get the sensing
error, is to find the receptor correlation time $\tau_c$, which
depends on the scenario by which the ligand finds the receptor.

\subsection{Expression of Berg and Purcell}
To obtain the receptor correlation time $\tau_c$, Berg and Purcell
assumed that the ligand binds the receptor in a Markovian fashion,
which means that $\tau_c$ is given by
\begin{equation}
\tau_c = \frac{1}{k_f c + k_b},
\elabel{BPtau_n}
\end{equation}
where $k_f$ is the ligand-receptor binding rate and $k_b$ is the
unbinding rate. Berg and Purcell described the binding site as a
circular patch on the membrane, with patch radius $s$. To get the
forward rate $k_f$, they assumed $k_f$ is given by the
diffusion-limited binding rate $k_D$, but with the cross section
$s$ renormalized by the sticking probability. For the binding of a
ligand to a membrane patch, $k_f = k_D = 4 D s$. We will consider the
binding of ligand to a spherical receptor protein with
ligand-receptor cross section $\sigma$, in which case $k_f = k_D =
4\pi \sigma D$. To get the backward rate $k_b$, Berg and Purcell
exploited the detailed-balance condition $k_f c (1-p) = p k_b$, which
states that in steady state the net rate of binding equals the net
rate of unbinding.

Combining \erefstwo{error2}{BPtau_n} yields the following
expression of Berg and Purcell for the sensing error
\begin{equation}
\left(\frac{\delta c}{c}\right)^2_{BP} = \frac{2}{4 \pi \sigma D c
  (1-p)T}.
\elabel{BPerror}
\end{equation}
This expression can be understood intuitively: The factor $4 \pi \sigma D
c$ is the rate at which ligand molecules arrive at the receptor,
$1-p$ is the probability that the receptor is free, and hence $4 \pi \sigma D
c (1-p)$ is the count rate at which the receptor binds the ligand
molecules; $4 \pi \sigma D
c (1-p)$  multiplied with $T$ is thus the total number of counts in
the integration time $T$. Indeed, this expression states that the
fractional error $\delta c / c$ decreases with the square root of the
number of counts, as we would expect intuitively.

While this expression makes sense intuitively, there are two
problems. First, receptor-ligand binding is, in general, not
Markovian. To illustrate this, imagine for the sake of the argument
that a ligand-bound receptor is surrounded by a uniform, equilibrium
distribution of ligand molecules. If the receptor-bound ligand
dissociates, then the other ligand molecules will still have the
equilibrium distribution. If it rebinds and then dissociates again, 
the other ligand molecules will again still have the equilibrium distribution. The
problem arises when a) the rebinding of the dissociated ligand molecule
is pre-empted by the binding of another ligand molecule; {\em and} b) if this
second molecule dissociates from the receptor before the first has
diffused into the bulk. If this happens, then the receptor and the dissociated
ligand molecule at contact are no longer surrounded by a uniform equilibrium
distribution of ligand molecules. Indeed, the process of binding
generates non-trivial spatio-temporal correlations between the
positions of the ligand molecules, which depend on the history of the
association and dissociation events. This turns
an association-dissociation reaction into a complicated non-Markovian,
many-body problem, which can, in general, not be solved analytically. 

The second problem of the analysis of Berg and Purcell is that not all
ligand-receptor association reactions are diffusion limited. Berg and
Purcell were fully aware of this, but they argued on p. 208 of Ref. \cite{berg1977} that if the ligand ``{\it doesn't stick on
  its first contact, it may very soon bump into the site again---and
  again. If these encounters occur with a time interval short compared
  to $\tau_b$ [the time a ligand is bound], their result is equivalent
  merely to a larger value of $\alpha$ [the sticking probability]. As
  we have no independent definition of the patch radius $s$, we may as
  well absorb the effective $\alpha$ into $s$.''} This argument,
however, does not take into account that when a ligand arrives at the
receptor for the first time and does not stick immediately, it
may also return to the bulk, after which another ligand molecule may
bind. Moreover, a ligand molecule that has just dissociated from the
receptor may either rapidly rebind the receptor, or diffuse away from
it into the bulk. It thus remained unclear how accurate the
expression of Berg and Purcell, \eref{BPerror}, is.

\subsection{Expression of Bialek and Setayeshgar}
Bialek and Setayesghar sought to generalize the result of Berg and
Purcell by explicitly taking into account the receptor-ligand binding
dynamics \cite{bialek2005}. They considered a model in which
the ligand molecules can diffuse, bind the receptor upon
contact with an intrinsic association rate $k_a$, and unbind
from the receptor with an intrinsic dissociation rate $k_d$. This
model is described by the following reaction-diffusion equations
\begin{align}
\frac{dn(t)}{dt}&=k_a c(x_0,t)(1-n(t))-k_d n(t) \elabel{BSRD1}\\
\frac{\partial c(x,t)}{\partial t}&=D\nabla^2 c(x,t)-\delta
(x-x_0)\frac{dn(t)}{dt},\elabel{BSRD2}
\end{align}
where $c(x,t)$ is the concentration of ligand at position $x$ at time
$t$ and $x_0$ is the position of the receptor. To solve these
equations, Bialek and Setayesghar invoked the fluctuation-dissipation relation,
leading to a linearization of \erefstwo{BSRD1}{BSRD2}.

The resulting expression for the sensing error is 
\begin{align}
\elabel{BSerror}
\left(\frac{\delta c}{c}\right)^2_{BS}=\frac{1}{\pi  \sigma D cT}+
\frac{2}{k_ac\br{1-p}T}.
\end{align}
The first term describes the contribution to the sensing error from
the stochastic transport of the ligand molecules to the receptor by
diffusion. The second term describes the contribution from the
intrinsic stochasticity of the binding kinetics of the receptor: Even
in the limit that $D\to\infty$, such that the ligand concentration is
uniform in space at all times, the ligand concentration can still not
be measured with infinite precision because the receptor
stochastically switches between the bound and unbound states, leading
to noise in the estimate of the receptor occupancy. This term is
absent in \eref{BPerror} since Berg and Purcell assume that the
binding reaction is fully diffusion limited, meaning that the
intrinsic rates $k_a$ and $k_d$ go to infinity. Indeed, for fast,
diffusion-limited reactions, this term can be much smaller than the
first one.  The rate $k_a$ is the rate of ligand-receptor binding,
given that receptor and ligand are in contact. Its maximum rate is
given by transition-state theory, which yields the rate $k_{\rm TST}$
in the absence of any recrossings of the dividing surface that
separates the bound from the unbound state
\cite{Chandler78,becker:2012ej}. It is given by $k_{\rm TST} = k_0
\exp[-\beta \Delta F]$, where $\Delta F$ is the free-energy barrier
separating the bound form the unbound state, and $k_0$ is a kinetic
prefactor. For spherical molecules that can bind in any orientation,
it is given by $k_0 = \pi \sigma^2 \avg{|v_{RL}|}$, where $v_{RL}$ is
the relative velocity of ligand and receptor.  For diffusion-limited
reactions, $\Delta F=0$, and $k_a^{\rm max} = k_0 = \pi \sigma^2
\avg{|v_{RL}|}$, which is typically much larger than the
diffusion-limited rate $k_D = 4 \pi \sigma D$.

More generally, the first term on the right-hand side 
\eref{BSerror} presents a noise floor that is solely due
to the stochastic transport of the ligand to the receptor by diffusion,
independent of the binding kinetics of the ligand after it has arrived at
 the receptor. The first term is thus considered to be the
fundamenetal sensing limit set by the physics of diffusion \cite{bialek2005},
and it can be compared with the expression of Berg and Purcell,
\eref{BPerror}. It is clear that the expression of Bialek
and Setayesghar and that of Berg and Purcell differ by a factor
$1/(2(1-p))$. This difference can have marked implications.
Although the Bialek-Setayeshgar expression predicts
that the uncertainty due to diffusion remains bounded even in the limit that
$p\to 1$, the Berg-Purcell expression suggests that it diverges
in this limit. Intuitively, we expect a dependence on $p$, because a
higher receptor occupancy at fixed $k_D$ should reduce the count rate.

\subsection{The expression of Kaizu and coworkers}
To elucidate the discrepancy between \erefstwo{BPerror}{BSerror},
Kaizu and coworkers rederived the expression for the sensing error
\cite{Kaizu:2014eb}. They considered exactly the same model as that of
Bialek and Setayesghar \cite{bialek2005}, but analyzed it using the
large body of work on reaction-diffusion systems, developed by Agmon,
Szabo and coworkers \cite{Agmon1990}. The goal is to obtain the
zero-frequency limit of the correlation function, $\widehat{C}(s=0)$,
from which the correlation time and hence the sensing error can be
obtained, see \eref{sigma_nT}. The correlation function of any binary
switching process is given by 
\begin{align}
\elabel{C_n}
C_n\br{\tau}=p^0_\ast\br{p_{\ast|\ast}\br{\tau}-p_\ast^0}
\end{align}
where $p^0_\ast\equiv\overline{n}$ is the equilibrium probability for the bound
state ($\ast$) and $p_{\ast|\ast}\brin{\tau}=\avgin{n(\tau)n(0)}/\overline{n}$
is the probability the receptor is bound at $t=\tau$ given it was bound at
$t=0$. To obtain the correlation function, we thus need
$p_{\ast|\ast}\brin{\tau}=1-\mathscr{S}_{\rm rev}\br{t|\ast}$, where $\mathscr{S}_{\rm
  rev}\br{t|\ast}$ is the probability  that the receptor is free at time $t$ given that it
  was occupied at time $t=0$.  It is given by the {\em exact} expression
\begin{align}
\elabel{surv_1}
\mathscr{S}_{\rm rev}\br{t|\ast}=k_d\int_0^t \sqbr{1-\mathscr{S}_{\rm
                rev}\br{t'|\ast}}\mathscr{S}_{\rm rad}\br{t-t'|\sigma}dt'.
\end{align}
The subscript ``rev'' denotes that a reversible reaction is
considered, meaning that in between $t=0$ and $t$
the receptor may bind and unbind ligand a number of times. The
probability that a receptor-ligand pair dissociates between $t'$ and
$t'+dt'$ to form an unbound pair at contact is $k_d\sqbrin{1-\mathscr{S}_{\rm
rev}\brin{t'|\ast}}dt'$, while the probability that the free
receptor with a ligand molecule at contact at time $t'$ is still
unbound at time $t>t'$ is $\mathscr{S}_{\rm rad}(t-t'|\sigma)$; the
subscript ``rad'' means that we now consider an {\it irreversible}
reaction ($k_d=0$), which can be obtained by solving the diffusion
equation using a ``radiation'' boundary condition
\cite{Agmon1990}. 

While \eref{surv_1} is exact, it cannot be solved analytically,
because, as discussed above, an association-dissociation reaction is a non-Markovian,
many-body problem. To solve \eref{surv_1}, Kaizu and coworkers made
the assumption that after each receptor-ligand dissociation event, the
other ligand molecules have the uniform, equilibrium
distribution. Mathematically, this assumption can be expressed as
\begin{align}
\elabel{surv_2}
\mathscr{S}_{\rm rad}\br{t|\sigma}=\mathscr{S}_{\rm rad}\br{t|\rm eq}S_{\rm
rad}\br{t|\sigma},
\end{align}
where $\mathscr{S}_{\rm rad}(t|{\rm eq})$ is the probability that a
receptor which initially is free and surrounded by an equilibrium
distribution of ligand molecules remains free until at least a later
time $t$, while $S_{\rm rad}(t|\sigma)$ is the probability that a free
receptor that is surrounded by only one {\em single} ligand
molecule, which initially is at contact, is still unbound at a later
time $t$. To solve \erefstwo{surv_1}{surv_2}, a relation between
$\mathscr{S}_{\rm rad}\br{t|\rm eq}$ and $S_{\rm
rad}\br{t|\sigma}$ is needed, which can be obtained from $\mathscr{S}_{\rm
rad}\br{t|\rm eq}= e^{-c \int_0^t k_{\rm rad}\br{t'}dt'}$
\cite{Rice1985} and the detailed-balance relation for the
time-dependent bimolecular rate constant $k_{\rm rad}(t) = k_a S_{\rm rad}(t|\sigma)$
\cite{Agmon1990}. 

With these relations, \erefstwo{surv_1}{surv_2} can be solved, which,
together with \erefstwo{sigma_nT}{C_n}, yields the following expression for the
 the sensing error \cite{Kaizu:2014eb}:
\begin{align}
\elabel{final}
\left(\frac{\delta c}{c}\right)^2_{KZ}=\frac{2}{4\pi\sigma D
c\br{1-p}T}+\frac{2}{k_ac\br{1-p}T}.
\end{align}
The second term is identical to that of Bialek and Setayesghar,
\eref{BSerror}. However, the first term, which constitutes the
fundamental limit, disagrees with the expression of Bialek and
Setaeysghar, but agrees with that of Berg and Purcell,
\eref{BPerror}. This suggests that the expression of Berg and Purcell
is indeed the most accurate expression for the fundamental sensing limit.

But it could of course be that both the analysis of Berg and Purcell
and that of Kaizu {\em et al.} are inaccurate. To investigate this,
Kaizu and coworkers performed particle-based simulations of the same
model studied by Bialek and Setayesghar and Kaizu {\em et al.}; to
test the expression of Berg and Purcell, the system was chosen to be
deep in the diffusion-limited regime. The
simulations were performed using Green's
Function Reaction Dynamics, which is an exact algorithm to simulate
reaction-diffusion systems at the particle level in time and space,
and hence does not rely on the approximation used to derive the
analytical result of Kaizu {\it et al.}
\cite{VanZon2005,VanZon2005b,Takahashi2010}. 
\fref{fig1} shows the results for the zero-frequency limit of the power spectrum,
$P_n (\omega \to 0) = 2 \sigma^2_n \tau_c$, which provides a test for
the receptor correlation time $\tau_c$ and hence the sensing error
(see \erefstwo{error}{sigma_nT}), because $\sigma^2_n =
\overline{n}(1-\overline{n})$. It is seen that the prediction of Kaizu
and coworkers agrees very well with the simulation results, in contrast to that of Bialek and
Setagesghar.  This shows that the expression of Kaizu {\em et
  al.} and hence that of Berg and Purcell, is the most accurate expression for
the sensing precision.

\begin{figure}[!ht]
\begin{center}
  \includegraphics[]{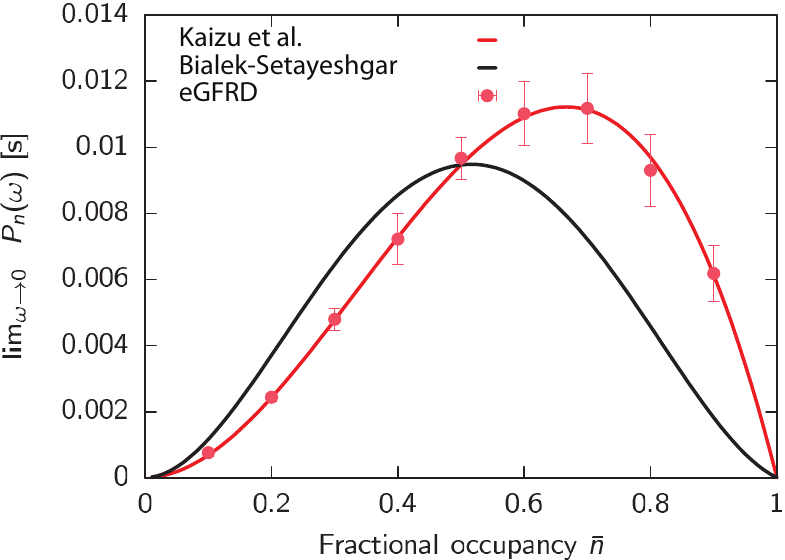}
\end{center}
\caption{\flabel{fig1}
 The zero-frequency limit of the power spectrum, $P_n (\omega \to 0) =
 2 \sigma^2_n \tau_c$ with $\sigma^2_n = \overline{n} (1-\overline{n})$,
as a function of the average receptor occupancy $\overline{n}$ for
$c=0.4\unit{\mu M}$; $\overline{n}$ is varied by changing $k_d$.
 It is
seen thatthe theoretical prediction of Kaizu et
al. \cite{Kaizu:2014eb} (red line)
agrees very well with the simulation results (red symbols), in
contrast to that of Bialek and Setayeshgar \cite{bialek2005} (black
line). 
Parameters: $D=1\unit{\mu m^2\,s^{-1}}$,
$\sigma=10\unit{nm}$, $L=1\unit{\mu m}$, $k_a=552\unit{\mu
  M^{-1}s^{-1}}$.}
\end{figure}

\subsection{Role of rebinding}
The question remains why the analysis of Kaizu {\em et al.} is so
accurate. The central assumption of \eref{surv_2} makes the
propensity for binding the next ligand independent of the history of
the previous binding events. In essence, it reduces
the non-Markovian many-body problem to a Markovian two-body problem,
which can be seen from the expression for the receptor correlation time 
$\tau_c=\brin{\sigma^2_n}^{-1}\widehat{C}_n\brin{s=0}$:
\begin{align}
\tau_c=\frac{1}{k_{\rm on}c+k_{\rm off}}.
\elabel{tau_n}
\end{align}
Here $k_{\rm on}$ and $k_{\rm off}$ are the renormalized association and
dissociation rates
\begin{align}
k_{\rm
on}&=\br{\frac{1}{k_a}+\frac{1}{k_D}}^{-1}=\frac{k_ak_D}{k_a+k_D},\elabel{kon}\\
k_{\rm off}&=\br{\frac{1}{k_d}+\frac{K_{\rm
eq}}{k_D}}^{-1}=\frac{k_dk_D}{k_a+k_D}.\elabel{koff}
\end{align}
with $K_{\rm eq}=k_a/k_d= k_{\rm on}/k_{\rm off}$ the equilibrium
constant and $k_D =  4\pi \sigma D$ the diffusion-limited rate constant---$k_D
= k_{\rm rad}(t\to \infty)$ for $k_a\to \infty$. \eref{tau_n} is the
expression for the correlation time of a receptor that switches in a memoryless fashion between the
bound and unbound states with switching rates $k_{\rm on}c$ and $k_{\rm off}$.

But why is \eref{surv_2} accurate? And what is the role of rebindings?
Do they not generate an algebraic tail in the correlation function? As
it turns out, these questions are intimately related. It is well known
that in an unbounded system, the correlation function exhibits an
algebraic tail because at long times the relaxation of the receptor
state is dominated by the slow diffusive transport of the ligand over
long distances
\cite{Popov2001,Gopich2002}. However, we typically expect the space to
be bounded, both for the binding of ligand to a receptor inside the
cell and to a receptor at the cell surface. Indeed, the simulations of
\cite{Kaizu:2014eb} were performed in a finite box of cellular
dimensions, yielding exponential, not algebraic, decay at long times. We then expect
\eref{surv_2} to become accurate. More specifically,
assumption
\eref{surv_2} breaks down when a) the rebinding of a dissociated
particle is pre-empted by the binding of another particle from the
bulk; {\em and} b) if this second particle dissociates from the
receptor before the first has equilibrated by diffusing into the
bulk. However, the time a ligand molecule spends near the receptor is
typically much shorter than the time for molecules to arrive from the
bulk at biologically relevant concentrations, which means that the
probability of rebinding interference is very small, and condition a)
is not met. Because biologically relevant concentrations are low, also
the dissociation rates are typically low, which means that also
condition b) is not met. The likelihood that both conditions are met,
necessary for the analysis to break down, is thus very small
\cite{Kaizu:2014eb}. 

Because rebindings are so much faster than bulk arrivals, they can be
integrated out \cite{vanzon2006,Kaizu:2014eb,Mugler:2013wx}. The
probability that a particle that has just dissociated from the
receptor will rebind the receptor rather than diffuse away into the
bulk is $p_{\rm reb}=1-S_{\rm rad}(\infty|\sigma) = k_a /
(k_a+k_D)$. The mean number of rounds of rebinding and dissociation
before the molecules diffuse into the bulk is then $N_{\rm reb} =
k_a/k_D$, which rescales the effective dissociation rate: $k_{\rm off} = k_d /
(N_{\rm reb} + 1)=k_d k_D / (k_a+k_D)$. Similarly, a molecule that
arrives at the receptor from the bulk may either bind the receptor or
escape back into the bulk with probability $p_{\rm esc} = 1-p_{\rm
  reb}$; the mean number of rounds of escape and arrival before
binding is $N_{\rm esc} = 1/N_{\rm reb}$, which rescales the effective
association rate $k_{\rm on} = k_D / (1+N_{\rm esc}) = k_a k_D /
(k_a+k_D)$. These are indeed precisely the rates of
\erefstwo{kon}{koff}. 

This analysis also elucidates the role of rebinding in
sensing. 
 The probability of rebinding does not depend on the
concentration, and rebindings therefore do not provide information on the
concentration. 
They merely increase the receptor correlation time by
increasing the effective receptor on-time
 from $k_d^{-1}$ to $k_{\rm off}^{-1}
= k_d^{-1}/(1+N_{\rm reb})$. After $(1+N_{\rm reb})$ rounds of
dissociation and rebinding, the molecule escapes into the bulk, and
then another molecule will arrive at the receptor with rate $k_Dc$;
this molecule may return to the bulk or bind the receptor, such that a
new molecule will bind after a time $(k_{\rm
  on}c)^{-1}$ on average. Importantly, this molecule will bind in a memoryless
fashion and with a rate that depends on the concentration. This
binding event thus provides an independent concentration measurement. The mean waiting time in between independent binding
events is therefore $\tau_w = 1/k_{\rm off}+1/(k_{\rm on}c)$, which
allows us to rewrite \eref{final} in a form that we would expect
intuitively:
\begin{align}
\elabel{BPerror2}
 \left(\frac{\delta c}{ c}\right)_{KZ}^2 = \frac{2}{k_{\rm on} c (1-p)
     T}=\frac{2\tau_w }{T}.
\end{align}
 Indeed, the sensing error $\delta c / c$
decreases with the square root of the number of independent measurements
$T/\tau_w$ during the integration time $T$.

Lastly, why does the expression of Bialek and Setayesghar,
\eref{BSerror}, miss the factor $1-p$ in the diffusion term? We
believe that this is because by invoking the fluctuation-dissipation
theorem, Bialek and Setayesghar effectively linearize the
reaction-diffusion problem, thereby ignoring correlations between the
state of the receptor and the local ligand concentration. 
This idea is
supported by the analysis of Berezhkovskii and Szabo
\cite{Berezhkovskii:2013hq}, who recently derived an expression for
the accuracy of sensing via multiple receptors on a sphere, ignoring
spatio-temporal correlations between the states of the respective
receptor molecules and the ligand concentration. Interestingly, in the
limit that the number of receptors on the sphere goes to infinity,
then, by replacing the radius of the cell with the cross section of
the receptor, their expression becomes identical to that of Bialek and
Setayesghar, see also \eref{BP_D} below; indeed, this expression does
not contain the factor $(1-p)$. This can be understood by noting that
in the limit that the number of receptors becomes large, there will
always be receptor molecules available for binding the ligand. But for
a single receptor, we have to take the binary character of the
receptor state into account.

\section{Can cells reach the Berg-Purcell limit?}
\label{sec:LinNet}
The work of Berg and Purcell and subsequent studies like those
discussed above
\cite{berg1977,Ueda:2007uq,bialek2005,levinepre2007,levineprl2008,levineprl2010,Skoge:2011gi,Skoge:2013fq,Kaizu:2014eb}
assume not only a given integration time $T$, but also that the
downstream signaling network averages the state of the receptor {\em
  uniformly} in time over this integration time $T$. It remained
unclear, however, how the signaling network determines the (effective)
integration time $T$, whether the network averages the signal
uniformly in time, and how this assumption affects the sensing
precision \cite{Govern2012}. It thus remained open whether signaling
networks can actually reach the Berg-Purcell limit. 

To address these questions, the authors of Ref. \cite{Govern2012}
considered linear, but otherwise arbitrary signaling network. For
deterministic networks of this type, the output $X(T_o)$ at time $T_o$ can be written as 
\begin{align}
X(t) = \int_{-\infty}^{T_o} \chi(T_o-t^\prime) RL(t^\prime) dt^\prime,
\end{align}
where $\chi(t-t^\prime)$ is the response function of the network and
$RL(t)$ is the stochastic receptor signal. To compare to previous
results, the authors assumed that at $t=0$ the environment changes
instantaneously and that the receptors and hence $RL(t)$ immediately
adjust, so that $RL(t)$ is stationary for $0<t<T_o$, with fluctuations
that decay exponentially with correlation time $\tau_c$
\cite{Kaizu:2014eb}. Moreover, they assumed that either: (1)
$\chi(T_o-t)=0$ for $t<0$, which corresponds to a scenario where the
response time $\tau_r$ of the network is shorter than $T_o$, or,
equivalently, the network reaches steady state by the time $T_o$; or
(2) $RL(t)=0$ for $t<0$, which corresponds to a scenario in which the
cell is initially in a basal state. In both cases, $X(t) =
\int_{0}^{T_o} \chi(T_o-t^\prime) RL(t^\prime) dt^\prime$. When
neither $\chi(T_o-t)$ nor $RL(t)$ are zero for $t<0$, then previous
states of the environment influence the state of the network at $T_o$,
which can either be a source of noise, or a source of information if
the environments are correlated.

\begin{figure*}[t]
\includegraphics[width=17cm]{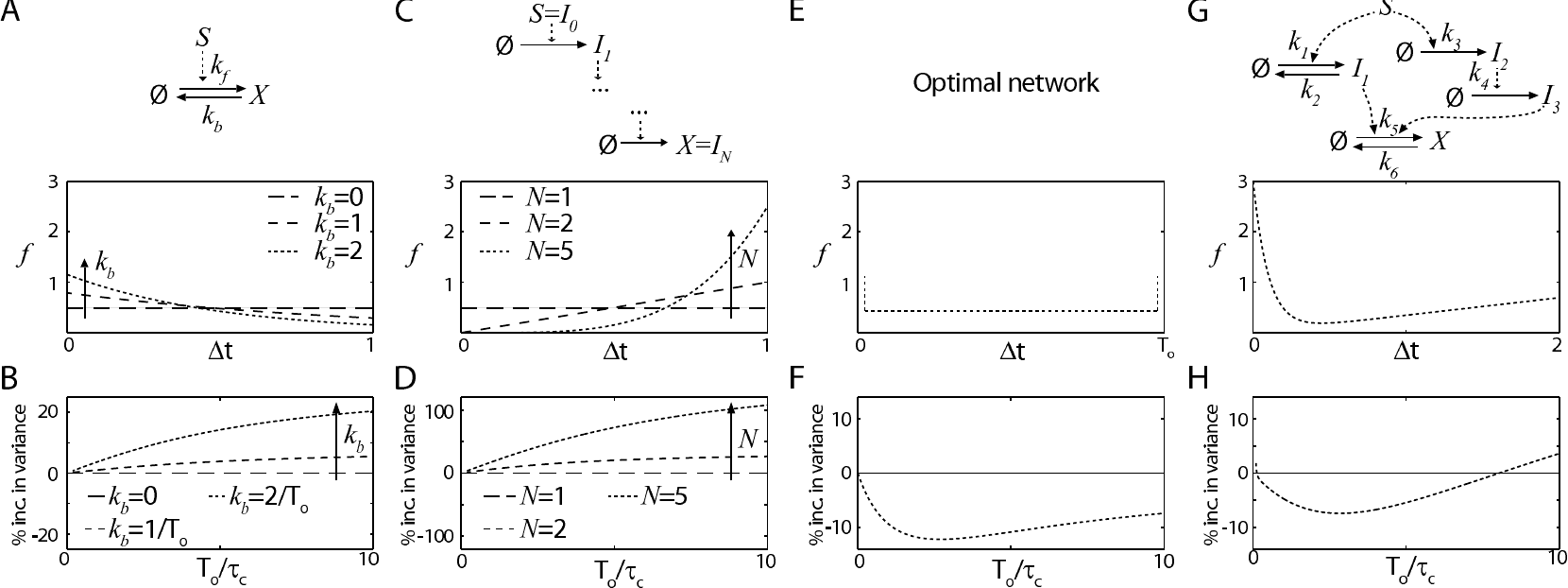}
\caption{ Extracting information from noisy input signals with linear
  signaling networks.  (A,C,E,G) The weighting functions corresponding
  to different signaling networks are not uniform. (B,D,F,H) The
  ability of a signaling network to measure ligand concentration
  depends on its weighting function.  The typical error (variance) in
  the estimate of ligand concentration is plotted as a percentage
  increase over the error of an estimate based on uniform weighting,
  assumed in the Berg-Purcell limit (Eq. 1 with $T=T_o$).  (A)
  Reversible, one-level cascades selectively amplify late
  ($t=T_{\rm{o}}$) values of the signal, (B) leading to worse
  performance than the uniform average.  (C) Irreversible, $N$-level
  cascades amplify early ($t=0$) values of the signal, (D) leading to
  worse performance than the uniform average. (E) The optimal
  weighting function, given in Ref. \cite{Govern2012}, averages the
  signal, selectively amplifying less correlated values.  The delta
  functions are truncated for illustration. (F) The optimal weighting
  function outperforms the uniform average. (G) A signaling network
  consisting of two branches, which selectively amplify late
  ($t=T_{\rm{o}}$) (left branch) and early ($t=0$) (right branch)
  values of the signal, approximates the optimal weighting function
  ($k_1 = 3.1$; $k_2 = 10$; $k_3 = 1$; $k_4 = 0.35$; $k_5 = 1$; $k_6
  \gg 1$; $T_{\rm{o}}=2$).  (H) The network in (F) can outperform the
  uniform average ($\tau_{\rm c}$ varies for fixed $T_{\rm{o}}=2$).
  \flabel{fig2} }
\end{figure*}

The idea is that the cell
infers the ligand concentration from the output $X(T_0)$ and by
inverting the input-output relation $\overline{X}(c)$. Using error
propagation, the error in the estimate of the concentration is then
given by
\begin{align}
\left(\frac{\delta c}{c}\right)_X^2 &= \frac{1}{c^2}
\left(\frac{dc}{d\overline{X}}\right)^2 \sigma^2_{X(T_0)}.
\elabel{Xerror}
\end{align}

The authors of Ref. \cite{Govern2012} then studied different signaling
architectures, shown in \fref{fig2}. Clearly, these networks do not, in
general, average the receptor signal uniformly in time; instead, they
have non-uniform weighting functions (\fref{fig2}A,C,E,G). They weigh
receptor signals in the past with a response function that depends on
both the lifetime of the signaling molecules and on the topology of
the signaling network. One-layer networks consisting of a single
reversible reaction give most weight to the most recent signal value
(left-most column), while multi-level cascades consisting of
irreversible reactions give more weight to signal values more in the distant
past (second column).  This concept can be generalized to arbitrarily
large signaling networks. Multilevel {\em reversible} cascades
have weighting functions that peak at some finite time
in the past, balancing the down-weighting of the signal
from the distant past due to the reverse reactions, with the
down-weighting of the signal from the recent past resulting
from the multilevel character of the network. Linear combinations of the weighting functions
for reversible and irreversible cascades can be achieved
with multiple cascades that are activated by the input in
parallel and which independently activate the same effector
molecule. Clearly, signaling
networks allow for very diverse weighting functions.

This idea can be exploited to improve the accuracy of sensing, as
shown in the right two columns of \fref{fig2}. A network with a
feedforward topology that
combines a fast reversible cascade with a slow irreversible cascade cannot only reach the
Berg-Purcell limit, but even beat it by $12\%$, when the observation
time $T_o$ is on the order of the receptor correlation time
$\tau_c$. The reason is that this network selectively amplifies the
more recent signal values and those further back in the past.
This is beneficial, because these signal values are less
correlated. 

While the data processing inequality suggests that it is advantageous
to limit the number of nodes in a signaling network to minimize the
effect of intrinsic noise, this study shows that there can be a
competing effect in favor of increasing the number of nodes: better
removal of extrinsic noise. Additional nodes make it possible to
sculpt the weighting function for averaging the incoming signal,
allowing signaling networks to reach and even exceed the Berg-Purcell
limit.

\section{Fundamental sensing limit of equilibrium systems}
Signaling networks are stochastic in nature, which means that
while they may remove the extrinsic noise in the input signal, they
will also add their own intrinsic noise to the transmitted signal.  Most studies on the
accuracy of sensing have ignored this intrinsic noise of the signaling
network
\cite{berg1977,Ueda:2007uq,bialek2005,levinepre2007,
  levineprl2008,wingreen2009,endres2010,levineprl2010,Govern2012,Skoge:2011gi,Skoge:2013fq,Kaizu:2014eb,Lang:2014ir}. They
essentially assume that the intrinsic noise can be made arbitrarily
small and that the extrinsic noise in the receptor signal can be
filtered with arbitrary precision by simply integrating the receptor
signal for longer.  However, the extrinsic and intrinsic noise are not
generally independent \cite{TanaseNicola2006}. This raises the
question whether the extrinsic and intrinsic noise can be lowered
simultaneously, and if so,  what resources would be required to achieve
this.

To address these questions, the authors of \cite{Govern:2014ez} first
studied equilibrium networks that are not driven out of thermodynamic
equilibrium via the turnover of fuel. Inspired by one component
signaling networks \cite{Ulrich:2005ys}, they started with the
simplest possible equilibrium network, consisting of
cytoplasmic readout molecules ${\rm X}$ that directly bind ligand-free
receptors ${\rm R}$: ${\rm R} + {\rm L} \rightleftharpoons {\rm RL}$,
${\rm R} + {\rm X} \overset{k_{\rm f}} {\underset{k_{\rm r}}\rightleftharpoons}
{\rm RX}$.  The linearized deviation $\delta x(t)=X(t)-\overline{X}$
of the copy number $X(t)$ from its steady-state value $\overline{X}$ is
\begin{equation}
\delta x(t) = \int_{-\infty}^t \chi(t-t^\prime) \left[\gamma RL(t^\prime) +
  \eta(t^\prime)\right],
\elabel{EqX}
\end{equation}
where $\chi(t-t^\prime) = e^{-(t-t^\prime) / \tau_I}$ is the response
  function with $\tau_I=1/(k_f (\overline{X}+\overline{R})+k_r)$ the
  integration time,  $\gamma = k_{\rm f} \overline{X}$, $RL(t)$ is the
input signal and $\eta$ describes the intrinsic noise of the
signaling network, set by the rate constants and copy numbers. 

The sensing error can be computed, as before, via \eref{Xerror}.
Here, the variance $\sigma^2_x= \langle \delta x \rangle^2$ can be
decomposed into the sum of the extrinsic noise $\sigma_{{\rm ex},x}^2
\equiv \gamma^2 K_{\delta RL,\delta RL}$ and the intrinsic noise
$\sigma_{{\rm in},x}^2 \equiv \gamma K_{\delta RL,\eta}+ K_{\eta
  ,\eta}$, where $K_{A,B} = \int_{-\infty}^{t} \int_{-\infty}^{t}
e^{-(t-t_1')/\tau_{\rm I}} C_{A,B}(t_1',t_2') e^{-(t-t_2')/\tau_{\rm
    I}} dt_1 dt_2$ with the correlation function $C_{AB}(t_1,t_2) =
\langle A(t_1) B(t_2) \rangle$. This decomposition is not unique, but
in this form the extrinsic noise term features a canonical temporal
average of the input (receptor) fluctuations \cite{Paulsson:2004dh,
  Shibata:2005th, TanaseNicola2006}, which can be made arbitrarily
small by increasing the effective integration time of the
network. However, the authors of Ref. \cite{Govern:2014ez} found that
when doing so in a system with $R_T$ receptors would reduce the total
sensing error below $4/R_T$, the intrinsic
noise would inevitably rise. The network faces a fundamental trade-off
between the removal of extrinsic and intrinsic noise.

Signaling networks are usually far more complicated than one
consisting of a single readout species, and as discussed in the
previous section, additional network layers can reduce the sensing
error \cite{Govern2012}. This raises the question whether a more
complicated equilibrium network can overcome the limit set by the
number of receptors. Searching over all possible network topologies to
address this question is impossible. However, equilibrium systems are
fundamentally bounded by the laws of equilibrium thermodynamics,
regardless of their topology. Indeed, starting from the
grand-canonical partition function, one can show that for {\em any}
equilibrium network the gain $d\overline{X}/d\mu$, with $\mu=\mu^0+kT
\log (c)$ the chemical potential of the ligand, is given by the
co-variance $\sigma^2_{X,RL}$ between $X$ and $RL$, because RL (or, in
general, the
complex containing the ligand) is the species conjugate to the
chemical potential. This means that these systems face a trade-off
between gain (sensitivity) and noise: increasing the gain inevitably
increases the noise. This has marked implications: using
$d\overline{X}/d\mu = \sigma^2_{X,RL}$ and \eref{Xerror}, we find that
the sensing error based on the readout X is $\left(\delta c /
  c\right)_X^2 = \sigma^2_X / (\sigma^2_{X,RL})^2$ , while if the
receptors themselves are taken as the readout, the sensing error is
$\left(\delta c / c\right)_{RL}^2 = 1/\sigma^2_{RL}$. From this it
follows that
\begin{equation}
\elabel{Eqerror}
\left( \frac{\delta c}{c} \right)_{X}^2 =  \frac{ \sigma_{X}^2 \sigma_{RL}^2  }{  \left( \sigma_{X, RL}^2 \right)^2  } \left( \frac{\delta c}{c} \right
)_{RL}^2 \ge \left( \frac{\delta c}{c} \right)_{RL}^2\geq \frac{4}{R_T^2}.
\end{equation}
Here the first equality inequality on the right-hand side follows from
the fact that $|\sigma_{X,RL}^2|/\sqrt{\sigma_X^2 \sigma_{RL}^2}$ is a
correlation coefficient, which is always less than 1 in magnitude.
The second inequality follows from the observation that for any
stochastic variable $0<Y<a$, $\sigma^2_Y \leq a^2 / 4$, meaning that
$\sigma^2_{RL} < R_T^2 / 4$. \eref{Eqerror} thus shows that in
equilibrium systems a downstream signaling network can never improve
the accuracy of sensing. The sensing precision 
is limited by the total number of receptors $R_T$, regardless of how
complicated the downstream network is, or how many protein copies are
devoted to making it.

What is the origin of the sensing limit in equilibrium sensing
systems? These systems transduce the signal by harvesting the energy of ligand binding: this
energy is used to boot off the downstream signaling molecules from the
receptor. However, detailed balance, by putting a constraint on the
binding affinities of receptor-readout and receptor-ligand binding,
then dictates that receptor-readout binding also influences
receptor-ligand binding, thus perturbing the future signal. Indeed, the
trade-offs faced by equilibrium networks are all different
manifestations of their time-reversibility.  The only way for a
time-reversible system to ``integrate'' the past is for it to
integrate and hence perturb the future. Concomitantly, in a time
reversible system, there is no sense of ``upstream" and
``downstream'', concepts which rely on a direction of time
\cite{Feng2008}; $RL$ is as much a readout of $X$,
as the other way around.  While in equilibrium systems the readout
encodes the receptor state, the readout is not a stable memory that
is decoupled from changes in the receptor state: a change in the state
of the readout, induced by readout-receptor (un)binding, influences the
future receptor state.  This introduces cross-correlations between the
intrinsic fluctuations in the activation of the readout, modeled by
$\eta(t)$ in \eref{EqX}, and the extrinsic fluctuations in the input
$RL(t)$: $K_{RL,\eta}\neq 0$. It is these cross-correlations,
which ultimately arise from time reversibility, that lead, in
these equilibrium systems,
to a fundamental tradeoff between the removal of extrinsic and
intrinsic noise and between increasing the gain and suppressing the noise.

\section{Sensing in non-equilibrium systems}

To beat the sensing limit of equilibrium systems, energy and the
receptor need to be employed differently. Rather than using the energy
of ligand binding to change the state of the readout, the system
should use fuel. This makes it possible change the readout via chemical modification, with the receptor catalyzing the
modification reaction: ${\rm RL} + {\rm X} \to {\rm RL} + {\rm X}^*$. This decouples
receptor-ligand binding from receptor-readout binding: the activation
of the readout does not influence the future receptor signal, while,
conversely, a change in the receptor state does not affect the
stability of the readout.  Each readout molecule that has interacted
with the receptor provides a stable memory; collectively, the readout
molecules encode the history of the receptor state. This enables the
mechanism of time integration, in which the trade-off between noise and
sensitivity is broken, and the extrinsic and intrinsic noise can be
reduced simultaneously \cite{Govern:2014ez}.

Catalysts cannot change the chemical equilibrium of two reactions that
are the microscopic reverse of each other.  To make the average
state of the readout dependent on the average receptor occupancy, the
activation reaction ${\rm RL}+{\rm X} \to {\rm RL} + {\rm X}^*$ must
therefore be coupled to a reaction
that is not its microscopic reverse, and the system must be driven
out of equilibrium. This is precisely the canonical signaling
motif of a receptor driving a push-pull network. In such a network the
receptor itself or the enzyme associated with it, like CheA in {\it
  E. coli} chemotaxis, catalyzes the activation of a readout protein X
via chemical modification, {\it i.e.} the phosphorylation of the
messenger protein CheY; active readout molecules ${\rm X}^*$ can then decay
spontaneously or be deactivated by an enzyme, like the phosphatase
CheZ in {\it E. coli}, via a reaction that is not the microscopic
reverse of the activation reaction. Typically, the activation via
chemical modification is coupled to fuel turnover, while deactivation
is not; in {\em E. coli chemotaxis}, for example, phosphorylation of
CheY is fueled by ATP hydrolysis: ${\rm CheA} + {\rm ATP} + {\rm CheY}
\to {\rm CheA} + {\rm ADP} + {\rm CheY_p}$, while dephosphorylation is
not: ${\rm CheZ} + {\rm CheY_p} \to {\rm CheZ}+{\rm CheY}+{\rm
  Pi}$. Another classical example is MAPK signaling, where activation
of MAPK is driven by ATP hydrolysis, while deactivation is not (even
though it is typically catalyzed by a phosphatase). In all
these systems, ATP hydrolysis is used to drive the readout molecule to
a high energy state, the active phosphorylated state, which then
relaxes back to the inactive dephosphorylated state via another
pathway, setting up a cycle in state space leading to energy
dissipation.

\subsection{The sensing error}
To derive the fundamental resources required for sensing, it is
instructive to view the downstream system as a
device that samples the state of the receptor discretely
\cite{Govern:2014ef}. The activation reaction ${\rm RL}+{\rm X}+{\rm ATP} \overset{k_{\rm
    f}} \to {\rm RL} + {\rm X}^*+{\rm ADP}$ (assumed to be fueled by ATP hydrolysis) generates samples of the ligand-binding state of
the receptor by storing the receptor state in the stable modification
states of the readout molecules.  We
expect that if there are $N$ receptor-readout interactions, then the
cell has $N$ samples of the receptor state and the error in the
concentration estimate, $\delta c/c$, is reduced by a factor of
$\sqrt{N}$. However, to derive the effective number of samples, we have to
consider not only the creation of samples, but also their decay and
reliability. The decay reaction ${\rm X}^* \overset{k_{\rm r}} \to
{\rm X}$ is
equivalent to discarding or erasing samples. The microscopically
reverse reactions of these activation and deactivation reactions,
namely the
receptor-mediated deactivation ${\rm X^*} + {\rm RL} +{\rm ADP}
\overset{k_{\rm -f}}\to {\rm X} +
{\rm RL} + {\rm ATP}$ and the spontaneous (or phosphatase catalyzed) activation
${\rm X}\overset{k_{\rm -r}} \to {\rm X}^*$ independent of the receptor, generate
incorrect samples of the receptor state. Energy is needed to
break time-reversibility and to protect the coding.

How the receptor samples are generated, erased, and how they are
stored in the readout, determine the number of samples, their
independence, and their reliability, which together set the sensing
precision \cite{Govern:2014ef}:
\begin{align}
\elabel{twoterm}
\left( \frac{\delta c}{c} \right)^2 = \frac{1}{p (1-p)}
\frac{1}{\bar{N}_I} + \frac{1}{(1-p)^2} \frac{1}{\bar{N}_{\rm eff}}.
\end{align}
In this expression, obtained using the here accurate linear-noise
approximation \cite{TanaseNicola2006,Govern:2014ef}, $p$ is as
before the probability that a receptor is bound to ligand. The
quantity $\bar{N}_I$, discussed below, is the average number of
receptor samples that are independent out of a total of $\bar{N}_{\rm
  eff}$ samples.  The first term is the error on the concentration
estimate that would be expected on the basis of $\bar{N}_I$ perfect,
independent samples of the receptor state that can be unambiguously
identified (as in \eref{error2}). A second correction term arises,
however, because the cell cannot distinguish between those readout
molecules that have collided with an unbounded receptor since their
last dephosphorylation event, and those that have not.

\begin{figure}
\includegraphics[width=\columnwidth]{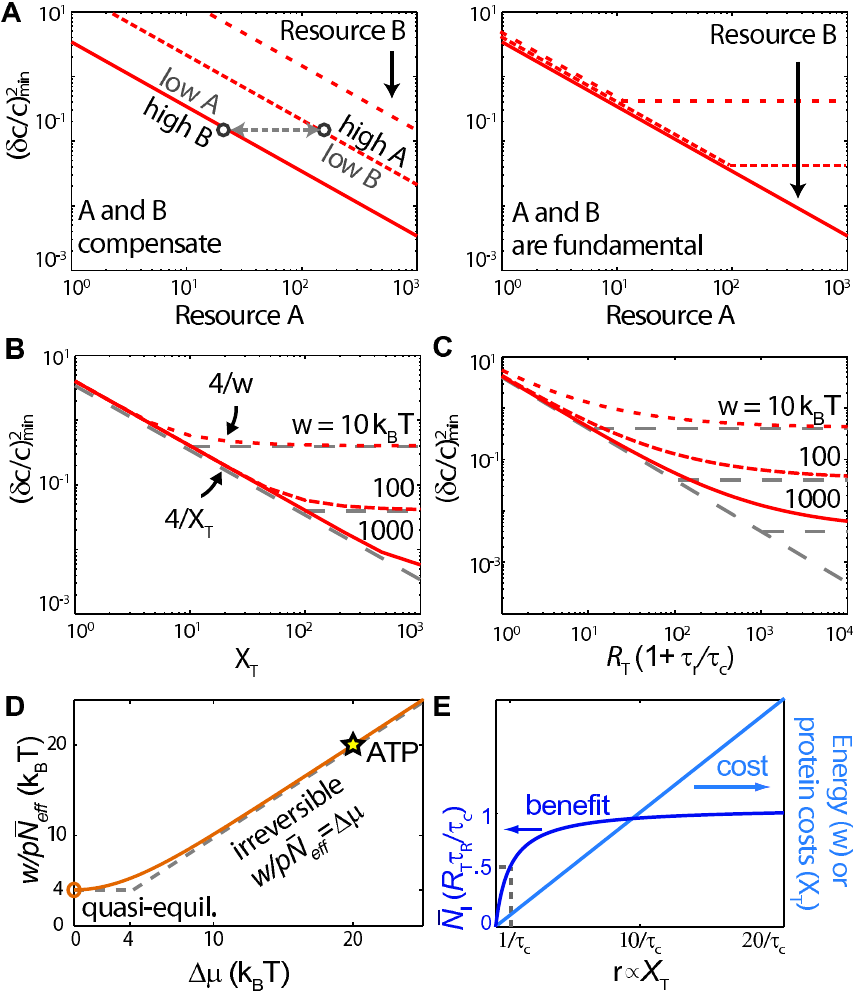}
\caption{
  \flabel{efficiency} Trade-offs in non-equilibrium sensing.   (A) When two resources A and B compensate each
other, one resource can always be decreased without affecting the sensing
error, by increasing the other resource; concomitantly, increasing a
resource will always reduce the sensing error.  When both resources
are instead fundamental, the sensing error is bounded by the limiting
resource and cannot be reduced by increasing the other.  (B,
C) The three classes time/receptor copies, copies of downstream
molecules, and energy are all required for sensing, with no trade-offs
among them. The minimum sensing error
obtained by minimizing \eref{twoterm} is plotted for different
combinations of (B) $X_T$ and $w$, and (C)
$R_T(1+\tau_r/\tau_c)$ and $w$.  The curves track the bound for the limiting
resource indicated by the grey lines, showing that the resources do
not compensate each other. The plot for the minimum sensing error as a
function of $R_T(1+\tau_r/\tau_c)$ and $X_T$ is identical to that of
(C) with $w$ replaced by $X_T$.   (D) The energy requirements for sensing.  In
the irreversible regime ($\Delta \mu \to \infty$), the work to
take one sample of a ligand-bound receptor, $w/(p\bar{N}_{\rm eff})$,
equals $\Delta \mu$, because each sample requires the turnover of one
fuel molecule, consuming $\Delta \mu$ of energy.  In the
quasi-equilibrium regime ($\Delta \mu \to 0$), each
effective sample of the bound receptor requires $4 {\rm k_BT}$, which
defines the fundamental lower bound on the energy requirement for
taking a sample.  When $\Delta \mu = 0$, the network is in equilibrium
and both $w$ and $\bar{N}$ are $0$. ATP hydrolysis provides $20 {\rm
  k_BT}$, showing that phosphorylation of read-out molecules makes
  it possible to store the receptor state reliably. The results are
obtained from\eref{neff} with $\Delta \mu_1 = \Delta \mu_2 =
\Delta \mu/2$. (E)
Sampling more than once per correlation time requires more resources,
while the benefit is marginal.  
As the sampling rate is increased
by increasing the readout copy number $X_T$, the number of independent
measurements $\bar{N}_I$ saturates at the Berg-Purcell limit $R_T
\tau_r/\tau_c$, but the energy and protein cost ($\propto X_T$) continue to rise.
}
\end{figure}

The number of independent measurements $\bar{N}_I$ can be expressed in
terms of collective variables that describe the
resource limitations of the cell
\begin{equation}
\elabel{neff}
\bar{N}_I =   \underbrace{\frac {1}{(1+2\tau_c/\Delta)}}_{f_I} \underbrace{\overbrace{\frac{\left( e^{\Delta \mu_1} -
        1 \right) \left( e^{\Delta \mu_2} - 1 \right)}{ e^{\Delta \mu}
      - 1}}^{q}\overbrace{\frac{\dot{n}
      \tau_r}{p}}^{\bar{N}} }_{\bar{N}_{\rm eff}}.
\end{equation}
This expression has a clear interpretation.  The relaxation time
$\tau_r$ is the effective integration time. The quantity $\dot{n}$ is
the flux of X across the cycle of activation by the receptor and
deactivation.  The product $\dot{n} \tau_r$ is thus the number of cycles of
read-out molecules involving collisions with ligand-bound receptor
molecules during the integration time $\tau_r$. The quantity $\dot{n}
\tau_r/p$ is the total number of read-out cycles involving collisions
with receptor molecules, be they ligand bound or not. It is thus the
total number of receptor samples taken during $\tau_r$, $\bar{N}$.

Not all of these samples are reliable. The effective number of samples
taken during $\tau_r$ is $\bar{N}_{\rm eff} = q \bar{N}$, where $0\leq
q\leq 1$ measures the quality of each sample. Here, $\Delta \mu_1$ and
$\Delta \mu_2$ are the average free-energy drops across the activation
and deactivation pathway respectively, in units of ${\rm k_ B T}$;
$\Delta \mu=\Delta \mu_1+\Delta \mu_2$ is the total free-energy drop
across the cycle, which is given by the free energy of the fuel
turnover, such as that of ATP hydrolysis. When $\Delta \mu=\Delta \mu_1=\Delta
\mu_2=0$, an active read-out molecule is as likely to be created by
the ligand-bound receptor as it is created spontaneously and there is
no coding and no sensing; indeed, in this limit, $q=0$ and
$\bar{N}_{\rm eff}=0$. In contrast, when $\Delta \mu_1, \Delta \mu_2
\to \infty$, $q \to 1$ and $\bar{N}_{\rm eff} \to \bar{N}$.

The factor $f_I$ denotes the fraction of samples that are
independent. It depends on the correlation time $\tau_c$ of
receptor-ligand binding and on the time interval $\Delta =  2 \tau_r/(\bar{N}_{\rm
  eff}/R_{T})$ between samples of the same
receptor. Samples farther apart are more independent.

\subsection{Fundamental resources and trade-offs}
\label{sec:NEQ_RES}
\erefstwo{twoterm}{neff} can be used to find the
resources that fundamentally limit sensing. A
fundamental resource or combination of resources is a (collective)
variable that when fixed, puts a lower bound on the sensing error, no
matter how the other variables are varied. It can be find via
constraint-based optimization, yielding \cite{Govern:2014ef}:
\begin{equation}
\left(\frac{\delta c}{c}\right)^2 \geq {\rm
  MAX}\left(\frac{4}{R_T\tau_r/\tau_c},\frac{4}{X_T},\frac{4}{\dot{w}\tau_r}\right).
\elabel{error_min}
\end{equation}
This expression identifies three fundamental resource classes, each
yielding a fundamental sensing limit:
$R_T(1+\tau_r /\tau_c)$, which for the relevant regime of time
integration $\tau_r>\tau_c$ is $R_T
\tau_r / \tau_c$, $X_T$, and $\dot{w} \tau_r$. These classes cannot compensate each other
in achieving a desired sensing precision, and hence do not trade-off
against each other.  The sensing precision is, like the weakest link
in a chain, bounded by the limiting resource, as illustrated in
\fref{efficiency}A-C. However, within each class, trade-offs are
possible. We now briefly discuss the fundamental resource classes and
their associated sensing limits.

{\em Receptors and their integration time, $R_T \tau_r / \tau_c$}. The
number of receptor samples increases with the number of readout
molecules, $X_T$. In fact, as $X_T \to \infty$, the spacing between
the samples $\Delta \to 0$ and the effective number of receptor
samples $\overline{N}_{\rm eff} \to \infty$; this is indeed the
Berg-Purcell mechanism of time integration. However, each receptor can
take an independent concentration measurement only every $2\tau_c$,
meaning that the number of independent measurements taken during the
integration time $\tau_r$ is, per receptor, $\tau_r /
\tau_c$ (the disappearance of the factor 2 is due to the fact that the
deactivation of X increases the effective spacing between the samples,
see \cite{Govern:2014ef}). Assuming that the receptors bind
independently (but see section \ref{sec:SpatTemp}), the total number
of independent concentration measurements, $\overline{N}_I$, taken
during $\tau_r$, is then limited by $R_T \tau_r /
\tau_c$, no matter how large $X_T$ is (\fref{efficiency}E).  This
yields the sensing limit of Berg and Purcell, $\left( \delta c /
  c\right)^2 \geq 4 / (R_T \tau_r / \tau_c)$, recognizing that the receptors are assumed to bind
independently, and $p(1-p)\leq 0.25$ ({\it cf.} \eref{error2}). While
the product $R_T \tau_r /\tau_c$ is fundamental, $R_T$ and $\tau_r$
are not: the error is determined by the total number of independent
concentration measurements, and it does not matter whether these
measurements are performed by many receptors over a short integration
time or by one receptor over a long integration time.

{\em The number of readout molecules, $X_T$}. Each
concentration measurement needs to be stored in the chemical
modification state of a readout molecule, and $X_T$ limits the maximum
number of measurements that can be stored. Consequently, no matter how
many receptors the cell has, or how much time it uses to integrate the
receptor state, the sensing error is fundamentally limited by the pool
of readout molecules, $\left(\delta c / c\right)^2 \geq 4 /X_T$. 

{\em Energy, $\dot{w}\tau_r$, during the integration time}.  The
power, the rate at which the fuel molecules do work, is
$\dot{w}=\dot{n}\Delta \mu$, and the total work performed during the
integration time is $w\equiv \dot{w} \tau_r$. This work is spent on
taking samples of receptor molecules that are bound to ligand, because
only they can modify $X$. The total number of effective samples of
ligand-bound receptors obtained during $\tau_r$, is
$p\overline{N}_{\rm eff}$.  Hence, the work needed to take one
effective sample of a ligand-bound receptor is $w/(p\bar{N}_{\rm
  eff})= \Delta \mu / q$ (see \eref{neff}). \fref{efficiency}D shows
this quantity as a function of $\Delta \mu$. In the limit that $\Delta
\mu \gg 4 k_BT$, $w /(p\bar{N}_{\rm eff})=\Delta \mu$, because the
quality factor $q\to 1$; in this regime, each receptor state is
reliably encoded in the chemical modification state of the readout,
and increasing $\Delta \mu$ further increases increases the sampling
cost with no reward in accuracy.  In the opposite regime, $\Delta \mu
< 4 k_B T$, however, the quality of the samples, $q$, rapidly
decreases with decreasing $\Delta \mu$. In this regime, the system
must take multiple noisy receptor samples to give the same information
as one single perfect sample. In the limit $\Delta \mu \to 0$, the quality factor $q \to \Delta \mu / 4$
and the work to take one effective sample of a ligand-bound receptor
approaches its minimal value of $w/(p\overline{N}_{\rm eff})=\Delta
\mu / q = 4 kT$.  Substituting this in \eref{twoterm} yields another
bound on the sensing error: $\left( \delta c/c \right)^2 \geq
4/(\dot{w}\tau_r)$. The bound can be reached when $R_T \tau_r /
\tau_c$ and $X_T$ are not limiting, and $\Delta \mu \to 0$. This bound
shows that while the total work $w=\dot{w}\tau_r$ done during the
integration time $\tau_r$ is fundamental, the power $\dot{w}$ and
$\tau_r$ are not, leading to a trade-off between accuracy, speed and
power, as found in adaptation \cite{Tu2012}.

\subsection{Design principle of optimal resource allocation}
The observation that resources cannot compensate each other,
naturally yields the design principle of optimal resource allocation,
which states that in an optimally designed system, each
resource is equally limiting so that no resource is in excess and thus
wasted. Quantitatively, \eref{error_min} predicts that in an optimally designed
system
\begin{align}
R_T \tau_r / \tau_c \approx X_T \approx w.
\elabel{opt_sys}
\end{align}
In an optimal sensing system, the number of independent concentration
measurements $R_T\tau_r / \tau_c$ equals the number of readout
molecules $X_T$ that store these measurements and equals the work (in
units of $k_BT$) to create the samples. 
Interestingly, the authors  of Ref. \cite{Govern:2014ef} found that the
chemotaxis system of {\it E. coli} obeys the principle of optimal
resource allocation,  \eref{opt_sys}. This indicates that there is a
selective pressure on the optimal allocation of resources in cellular sensing.

\section{Discussion}
\subsection{Different sensing strategies encode and decode ligand
  information differently}
Cells use different sensing strategies, which differ in how they
process information about the ligand concentration.  The data
processing inequality \cite{Cover2012} guarantees for any network that
no readout $X$ can have more information about the ligand concentration
encoded in its time trace than the ligand-bound receptor $RL$ has in its
time-trace \cite{Govern:2014ez}: $I( X_{[0,T]}(t); \mu_L) \le I(
RL_{[0,T]}(t); \mu_L)$, where $I$ is the mutual information between
the arguments with $\mu_L$ the chemical potential of the ligand, and
$y_{[0,T]}(t)$ indicates the time trace of $y=X,RL$ from time 0 to
time $T$. Clearly, the accuracy of sensing for any network is bounded
by the amount of information that is in the time trace of the receptor
state. However, the different sensing strategies differ in how they
encode the ligand concentration in the receptor dynamics and in how they
decode the information that is in the receptor time trace.

For equilibrium networks, the data processing inequality guarantees that no
readout has more information about the ligand than the receptors at
any given time \cite{Govern:2014ez}: $I(X(T); \mu_L) \le I(RL(T);
\mu_L) \le \log_2 (R_{\rm T} + 1)$, and therefore the information in
the instantaneous level of the readout is bounded by the total number of
receptors $R_T$. This statement is the information-theoretic analogue of
\eref{Eqerror}.  The history of receptor states does contain more
information about the ligand concentration than the instantaneous
receptor state, but an equilibrium signaling network cannot exploit this:
its output contains no more information than the instantaneous
receptor state.

Cells that use the mechanism of time integration can exploit the
information that is the time trace of the receptor, and for these
networks $I(X(T);\mu_L)$ can be larger than $I(RL(T);\mu_L)$. These
cells estimate the ligand concentration from the average receptor
occupancy over an integration time, which, as we have seen in section \ref{sec:LinNet}, is determined by the
architecture of the readout system and the lifetime of the readout
molecules. It is quite clear that cells employ this mechanism of
time integration: the central motif of cell signaling in both
prokaryotes and eukaryotes, the push-pull network, implements time
averaging by storing the receptor state into stable chemical
modification states of the readout molecules, which, collectively,
encode the average receptor occupancy over the past
integration time. 

Another sensing strategy is maximum likelihood estimation
\cite{wingreen2009,mora2010,Lang:2014ir}. It estimates the ligand
concentration not from the average receptor occupancy over the
integration time $T$, as in the mechanism of time integration, but
rather from the mean duration of the unbound state of the receptor
$\tau_u$: $\hat{c}_{\rm MLE}=1/(\tau_u k_{\rm on})$. The sensing error
of this strategy for a single receptor is $\left(\delta c / c
\right)^2_{\rm MLE} = 1 / (k_{\rm on} c (1-p) T)$ \cite{wingreen2009},
which is half that of the mechanism of time integration, see
\eref{BPerror2}. The reason why this sensing strategy is more accurate
is that only the binding rate depends on the concentration, not the
unbinding rate. Hence, only the unbound interval provides information
on the concentration. In contrast, the mechanism of time integration
infers the concentration from the mean receptor occupancy, which
depends on both the unbound interval and the uninformative bound
interval.

How cells could actually implement the strategy of maximum-likelihood
estimation remains an open question. One possibility is that receptors
are internalized upon ligand binding, another that they bind ligand
only briefly and signal only transiently,  which could be achieved
via receptor adaptation or desensitization following ligand binding \cite{wingreen2009}. Another
intriguing possibility has recently been suggested by Lang {\it et al.}
\cite{Lang:2014ir}. It is inspired by the observation that many
receptors, such as receptor-tyrosine kinases and G-protein coupled
receptors, are chemically modified via fuel turnover
\cite{Lang:2014ir}. In this scheme, the cell estimates the ligand
concentration from the average receptor occupancy over an integration
time $T$, as in the canonical mechanism of time integration. However,
upon ligand binding, the receptor is driven via fuel turnover through
a non-equilibrium cycle of $m$ chemical modification steps, before it
can release and bind new ligand again. In the limit that the energy
drop over the cycle $\Delta \mu \to \infty$ and $m\to \infty$, the
sensing accuracy approaches the maximum-likelihood-estimation limit,
even though the concentration is inferred from the average receptor
occupancy. The reason is that in this limit the interval distribution
of the active receptor state becomes a delta function instead of an
exponential one as in the case of canonical time integration. This
eliminates the noise from the uninformative bound interval in
estimating the average receptor occupancy.

\subsection{The importance of spatio-temporal correlations}
\label{sec:SpatTemp}
Ultimately, the precision of sensing via a mechanism that relies on
integrating the receptor state, be it the canonical Berg-Purcell
scheme with Markovian active receptor states or the
maximum-likelihood scheme of Lang {\em et al.} with non-Markovian
active states \cite{Lang:2014ir}, is determined by the
number of receptors, the receptor correlation time, and how the
readout molecules sample the receptor molecules. The analysis of
Ref. \cite{Govern:2014ef} ignores any spatio-temporal correlations of
both the ligand molecules and the readout molecules. In this analysis,
the different receptor molecules bind the ligand molecules
independently, and the correlation time of the receptor cluster is
that of a single receptor molecule $\tau_c$. The total number of
independent concentration measurements in the integration time $T$ is
then the number of receptors $R_T$ times the number of independent
measurements per receptor, $T / \tau_c$, yielding the fundamental
limit $\left(\delta c/ c\right)^2 \geq 2 \tau_c/ (p(1-p)R_T
T)$. Importantly, because $\tau_c$ is independent of the number
of receptors, the sensing error decreases with the number of
receptors.  However, diffusion introduces spatio-temporal correlations
between the different ligand-receptor binding events
\cite{berg1977,bialek2005,levinepre2007,Berezhkovskii:2013hq}. Consequently,
the correlation time $\tau_N$ of $R_T$ receptors on a spherical cell
of radius ${\cal R}$ is not that of a single receptor molecule, but
is rather given by \cite{Berezhkovskii:2013hq}
\begin{align}
\tau_N = \frac{1}{k_a c + k_d} + \frac{k_a\left(k_a c + R_T
    k_d\right)}{4\pi D {\cal R}
  \left(k_a c +  k_d\right)^2}.
\elabel{tau_N}
\end{align}
As pointed out by Wang {\em et al.} \cite{levinepre2007}, the correlation
time $\tau_N$ increases with the number of receptors $R_T$ (and even
diverges for $R_T\to \infty$), which means that when $R_T$ is large
and/or the integration time $T$ is short, the mechanism of time
integration breaks down. In this regime the equilibrium sensing
strategy is superior, because it relies on
sensing the instantaneous receptor state
\cite{levinepre2007}. Using receptors that bind ligand
non-cooperatively as the readout, $\left(\delta c / c\right)^2_{RL} =
1/\sigma^2_{RL}=1/((p(1-p) R_T)$, which indeed decreases with $R_T$ \cite{levinepre2007,Govern:2014ez}.

When the integration time $T$ is longer than $\tau_N$, the
sensing error is given by \cite{Berezhkovskii:2013hq}
\begin{align}
\left(\frac{\delta c}{c}\right)^2 &=\frac{2\tau_N}{R_T T p(1-p)}\\
&=\frac{2}{R_T k_a c T}\left(1+\frac{k_a
    c}{k_d}\right)+\frac{1}{2\pi D{\cal R}cT}\left(1+\frac{k_a c}{R_T
    k_d}\right).
\end{align}
For large $R_T$ (but not so large that $\tau_N > T$), the sensing
error reduces to 
\begin{align}
\left(\frac{\delta c}{c}\right)^2&=\frac{1}{2\pi D{\cal R}c T}.
\elabel{BP_D}
\end{align}
This, apart from the factor $1-p$, is the classical result of Berg and
Purcell \cite{berg1977,bialek2005}. At sufficiently large $R_T$, the
sensing error is limited by diffusion, the size of the cell and the
integration time. It becomes independent of $R_T$, because the decrease
of the instantaneous error with $R_T$, $1/(R_T(p(1-p))$, is cancelled
by the increase of the correlation time with $R_T$.

Not only in the encoding of the ligand concentration in the receptor
dynamics, but also in the decoding of this information by the readout
system, spatio-temporal correlations can become important. Receptor
and readout molecules are often spatially partitioned, due {\it e.g.},
to the underlying cytoskeletal network or lipid rafts. Even in a
system that is spatially homogeneous on average, spatio-temporal
partitioning would occur, because of the finite speed of diffusion. We have
recently shown that this partitioning decreases the propagation of
noise, essentially because the activation of the different readout
molecules becomes less correlated \cite{Mugler:2013bx}. Whether there
exists an optimal diffusion constant of the readout molecules that
matches the correlation length and time of the receptors, which is set by the ligand diffusion and binding dynamics, is an
intriguing question for future work.

\subsection{Cooperative receptor activation}
One important aspect that we have not addressed so far is the role of
receptor cooperativity. It is now well established that receptors are
often activated cooperatively, with the most studied and best
characterized example being the receptor cluster of the {\it E. coli}
chemotaxis system. How does this affect the precision of sensing? This
depends (again) on the receptor correlation time
\cite{Skoge:2013fq}. Skoge {\it et al.} found that while cooperative
interactions between neighboring receptors can increase the gain,
which reduces the sensing error, they also increase the correlation
time, such that independent receptors are, in fact, optimal. As we now
know, equilibrium systems do not rely on time integration, and hence
do not suffer from a slowing down of the receptor dynamics. In fact,
we have shown that for all equilibrium systems in which the receptors
bind the ligand {\em non}-cooperatively, $\left(\delta c / c\right)_X^2
\geq 1/R_T$; hence, to reach the bound of \eref{Eqerror} for {\em all}
equilibrium networks, cooperative ligand binding is necessary
\cite{Govern:2014ez}. In \cite{Govern:2014ez} we show that cooperative
ligand binding makes it indeed possible to beat the non-cooperative
bound, but whether equilibrium sensing systems can actually reach the
bound of \eref{Eqerror} remains an open question.

\subsection{The role of energy in sensing}
It seems intuitively clear that fuel turnover can be used to enhance
the precision of sensing, but {\em how} it can be used is less
obvious. In the maximum-likelihood scheme of Lang {\em et al.} it is
used to make the interval-distribution of the active receptor state
more deterministic \cite{Lang:2014ir}. In the scheme of time
integration, fuel turnover is used to sample the receptor state
\cite{Mehta2012,Govern:2014ef}. 

The latter example seems tantalisingly related to the thermodynamics
of computation, formulated by Bennett and Landauer decades ago
\cite{Bennett:1982hi,Landauer1961}. In particular, the receptor state
appears to be copied into the chemical modification
states of readout molecules, which thereby acts as memory
elements for time integration
\cite{Mehta2012,Govern:2014ef}. Performing copy operations
  repeatedly using the same readout requires net work input, unless
  the correlation between the data bit (receptor) and the memory
  (readout), generated by the copy operation, is used to extract work
\cite{Ouldridge:2015vi}.  Indeed, the arguments of Landauer and
  Bennett \cite{Bennett:1982hi,Landauer1961} show that the
  minimal amount of work for a perfect copy cycle is $k_B T \ln(2)$.
But how does this bound apply to biochemical networks?

  To answer this question it is important to make a formal mapping
  between cellular sensing systems and copy operations.  As it
  turns out, cellular copy protocols differ fundamentally from ideal
  quasi-static protocols, such as those considered by Landauer and
  Bennett \cite{Bennett:1982hi,Landauer1961}.  Copying entails changing
  the state of the memory, which means that a thermodynamic driving
  force must be applied to the system. Thermodynamically optimal
  protocols increase the driving force slowly, such that the memory is
  slowly driven to its new state. In contrast, in cellular systems the
  thermodynamic driving force for the reactions that implement the
  copy process is typically constant, because the fuel molecules that
  drive these reactions are commonly present at constant concentration
  \cite{Ouldridge:2015vi}. As a result, cellular systems face a
  trade-off between cost and precision that is both qualitatively and
  quantitatively distinct from that required thermodynamically,
  regardless of parameters \cite{Ouldridge:2015vi}. They dissipate
  more to achieve the same accuracy. One of the most vivid
  manifestations of this difference concerns the Landauer limit
  itself. One of the surprising, but by now well-known, results of
  Bennett and Landauer was that quasi-static protocols make it
  possible to perform repeated copies with 100\% accuracy at only a
  finite energy cost of, indeed, $k_B T \ln(2)$ per copy. In contrast,
  cellular copy protocols can only reach 100\% accuracy when the cost
  diverges. For the purpose of sampling a noisy signal, however,
  perfect copies are not necessarily ideal. Indeed, as we have seen in
  section
\ref{sec:NEQ_RES}, the energetically most efficient approach to
record the receptor state is to take many noisy samples, which
together make up one effective sample. For the canonical push-pull
network considered here the minimal cost to take one effective
receptor sample is $2 k_B T$ on average if the receptor occupancy is $p=0.5$
\cite{Govern:2014ef}. For a
bi-functional kinase system, in which the kinase associated with the
receptor catalyzes the phosphorylation of the readout when the
receptor is bound to ligand, but dephosphorylation when the receptor
is not bound to ligand, this minimal cost
is even lower: $1k_BT$ \cite{Ouldridge:2015vi}.

\subsection{How resources determine the fundamental sensing limit: Trade-offs between equilibrium and non-equilibrium sensing}
Information processing devices require resources to be built and
run. Components are needed to construct the system, space is required
to accommodate the components and energy is required to make the
components and operate the system. These resources constrain the
design and performance of any device, and cellular sensing systems are
no exception. Making proteins is costly \cite{Dekel2005}. They also
take up valuable space: both the membrane and the cytoplasm are highly
crowded, with proteins occupying 25--75\% of the membrane area
\cite{Linden2012} and 20--30\% of the cytoplasmic volume
\cite{Ellis2001}. And many cellular signaling pathways, including
two-component systems in bacteria \cite{Stock:2000ve}, GTP-ase cycles
as in the Ras system \cite{Pylayeva-Gupta:2011kx}, phosphorylation
cycles as in MAPK cascades \cite{Chang:2001uq}, are driven out of
thermal equilibrium via the turnover of fuel.  Also the adaptation
system that allows {\em E. coli} to adapt to a wide range of
background concentrations is driven out of equilibrium
\cite{Tu2012}. However, cells also commonly employ equilibrium motifs,
such as protein binding and sequestration. Indeed, as we have
seen, sensing does not fundamentally require energy input
\cite{Govern:2014ez}. Equilibrium sensing systems can respond to
changes in the environment by harvesting the energy of ligand binding,
thereby capitalizing on the work that is performed by the environment
to change the ligand concentration. Also adaptation does not
fundamentally require energy consumption \cite{Endres}.

\begin{figure}[t]
\includegraphics[width=\columnwidth]{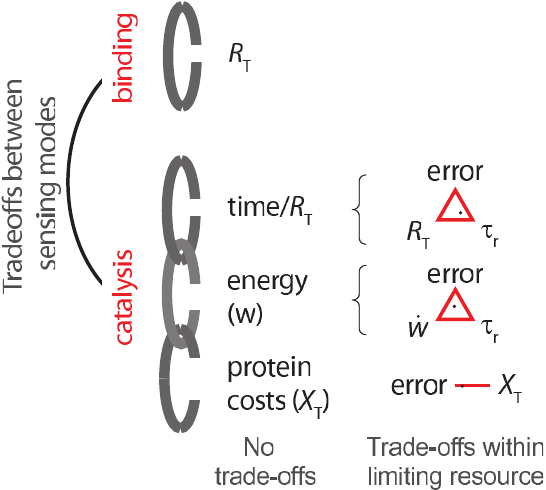}
\caption{
  \flabel{TradeOff} Cells face a fundamental trade-off between two
  modes of sensing, an equilibrium mode based on binding and
  sequestration and a non-equilibrium mode based on catalysis. These
  sensing strategies have different resource requirements.}
\end{figure}

When does the non-equilibrium sensing strategy outperform the
equilibrium one? This depends on the resources available to the cell,
as summarized in \fref{TradeOff}.  Comparing the bound for non-equilibrium
systems, \eref{error_min}, with that for equilibrium ones without
cooperative binding, $\left(\delta c / c\right)^2 \geq 1 / R_T$,
predicts that non-equilibrium systems can sense more accurately when
there is at least one readout molecule available per receptor, and
the amount of energy dissipated per receptor during the
integration time is at least $1k_B T$ \cite{Govern:2014ez}.

Interestingly, evolution may have toggled between equilibrium and
non-equilibrium sensing strategies. Bacteria employ both one- and
two-component signaling networks. One-component systems follow the equilibrium
strategy, consisting of adaptor proteins which can bind an upstream
ligand and a downstream effector. Two-component systems are
similar to the non-equilibrium push-pull system considered here, consisting of a
kinase (receptor) and its substrate. Intriguingly, some adaptor
proteins, like RocR, contain the same-ligand binding domain as the
kinase and the same effector-binding domain as the substrate of a
two-component system, i.e. NtrB-NtrC \cite{Ulrich:2005ys}. They could
thus transmit the same signal. Our results
suggest that these are alternative signaling strategies, selected
because of different resource selection pressures. It is tempting to
believe that when sensing precision is important, but space for
receptors on the membrane is limiting, non-equilibrium sensing becomes
essential, because it makes it possible to take more concentration
measurements per receptor. 

\section{Conclusion}
In this review we have focused on sensing concentrations that do not
vary on the timescale of the response of the system. While some
questions remain open, such as the importance of spatio-temporal
correlations in both ligand-receptor and receptor-readout binding,
this problem is by now fairly well understood. We understand how the
receptor correlation time depends on the diffusion and binding
kinetics of the ligand (although the question of the correlation time
of multiple receptors is, arguably, still open), how the effective
integration time depends on the lifetime of the readout molecules and
the architecture of the readout network, and how the precision of
sensing depends on the number of receptors, the number of readout
molecules, the receptor correlation time, the integration time, and
energy. We understand how combinations of resources impose fundamental
sensing limits and what this implies for the optimal design of
cellular sensing systems.

The challenge will be to make a similar leap for systems that do
not respond rapidly on the timescale of variations in the input
signal. For these systems, we have to take the dynamics of the input
signal into account. On this front, progress has been made in recent
years. We are now beginning to understand how in these systems
information transmission depends on the lifetime of the readout
molecules and on the topology of the readout network
\cite{tostevin2010,deRonde:2010hh,deRonde:2012fs,Aquino:2014co}, and what the
trade-off between energy dissipation and information processing is
\cite{Barato2013,Barato:2014ta,Horowitz:2014wb,Sartori:2014gz}. Yet,
many questions are still wide open: What is the performance measure
that best descibres the design logic of cellular sensing systems? Is
it the average sensing error, the instantaneous mutual information,
the information transmission rate \cite{tostevin2010}, or the learning
rate \cite{Barato:2014ta,Horowitz:2014wb}? What resource combinations
impose fundamental sensing limits? Also new questions arise: How
accurately can living cells predict the future input signal
\cite{Becker:2013uk}? And what are the thermodynamic costs of
cellular prediction \cite{Still:2012df,Becker:2013uk}? The physics of
sensing will remain a fascinating problem for many years to come.

\section{Acknowledgements}  This work is part of the research programme
of the Foundation for Fundamental Research on Matter (FOM), which is
part of the Netherlands Organisation for Scientific Research (NWO).

%\bibliography{sensingbib2}

\end{document}